\documentclass[11pt,a4paper]{article}

\usepackage[utf8]{inputenc}
\usepackage{a4wide}
\usepackage{amsmath}
\usepackage{graphicx}
\usepackage{authblk}
\usepackage{color}

\usepackage{multicol}

\begin{document}
\title{Transition State Theory Approach to Polymer Escape from a One Dimensional Potential Well}
\date{\today}

\author[1,2]{Harri M\"okk\"onen}
\author[1]{Timo Ikonen}
\author[1]{Tapio Ala-Nissila}
\author[1,2]{Hannes J\'onsson}

\affil[1]{\footnotesize{Department of Applied Physics and COMP CoE, Aalto University School of Science, P.O. Box 11000, FIN-00076 Aalto, Espoo, Finland \footnote{harri.mokkonen$@$aalto.fi}}}
\affil[2]{Faculty of Physical Sciences, University of Iceland, VR-III, 107 Reykjav{\'\i}k, Iceland}

\maketitle

\begin{abstract}

The rate of escape of an ideal bead-spring polymer in a symmetric double-well potential is calculated using  transition state theory (TST) and the results compared with direct dynamical simulations. The minimum energy path of the transitions becomes flat and the dynamics diffusive for long polymers  making the Kramers-Langer estimate poor. However, TST with dynamical corrections based on short time trajectories started at the transition state gives rate constant estimates that agree within a factor of two with the molecular dynamics simulations over a wide range of bead coupling constants and polymer lengths. The computational effort required by the TST approach does not depend on the escape rate and is much smaller than that required by molecular dynamics simulations.

\end{abstract}

%\pagebreak
%\tableofcontents

\pagebreak

%%%%%%%%%%%%%%%%%%%%
%  INTRODUCTION				  %
%%%%%%%%%%%%%%%%%%%%

\section{Introduction} \label{sec:intro}

The movement of a polymer from one metastable state to another (meta)stable state by crossing a free energy barrier through thermal activation is referred to as the polymer escape problem. The free energy barrier separating the states is most often of entropic origin due to geometric confinement. Relevant examples include polymer translocation \cite{Palyulin2014, Muthukumar2011}, where a polymer is crossing a membrane through a pore \cite{Kasianowicz1996}, or narrow  $\mu$m-scale channels with traps \cite{Han1999}. Recent experiments by Liu \textit{et al.} involve the escape of DNA molecules from an entropic cage \cite{Liu2014}. Similar translocation and escape processes are common in cell biology and have possible bioengineering applications, such as DNA sequencing \cite{Howorka2001} and biopolymer filtration \cite{Mikkelsen2011}. 

The polymer escape problem is a generic description of systems such as the ones described above. In the polymer escape problem there is a static barrier represented by an external potential energy function where the two minima are equal or the final state has lower energy than the initial state. Therefore, the escape problem is a combination of the classical thermal activation problem, where the free energy barrier is largely energetic \cite{Wigner1935, Kramers1940}, and the polymer translocation problem, where the free energy barrier is typically of entropic origin. 

The problem has been studied using Kramers' theory for polymers by Park and Sung, who evaluated the escape rate using lattice statistics of a discrete ideal polymer model \cite{Park1999}. Sebastian proposed a kink diffusion mechanism \cite{Sebastian2000} for long chains when one end of the polymer has moved over the barrier, while the other end is still in the initial state energy well. The kink corresponds to the beads in the region of the energy barrier and it moves along the chain as the polymer moves from one potential energy well to another. In Ref. \ref{Sebastian2000a}, Sebastian and Paul described the polymer escape of long chains as a two-step process. In the first step, the polymer is thermally activated to bring one end over energy barrier and into the final state well, and the second step involves diffusive motion of the kink as intermediate beads move in and out of the barrier region. A rate theory approach similar to Langer's \cite{Langer1969} multidimensional extension of Kramers' theory was proposed for activation \cite{Sebastian2000a}. More recently, Sebastian and Debnath studied the thermal activation mechanism for short chains \cite{Sebastian2006} and simulated kink diffusion for one and three dimensional systems \cite{Sebastian2010}. Lee and Sung studied the polymer escape in a symmetric external potential well and proposed a rate theory approach to predict the rate for linear \cite{Lee2001} and for star polymers \cite{Lee2001a}. They also found that for linear polymers the stretched kink solution is the dominant escape mechanism for chains longer than a certain crossover length $N_C$. Below $N_C$ the polymer crosses the barrier in a coiled form while polymers that are longer than $N_C$ are stretched during the transition, analogous to instantons in quantum mechanical tunnelling of one particle. Paul \cite{Paul2005} has studied polymer escape of star polymers in a system mimicking experiments carried out by Han \textit{et al.}  \cite{Han1999}.

The studies mentioned above only considered the escape of flexible, ideal polymers with zero bond length at zero temperature. The semiflexible case has been studied by Kraikivski \cite{Kraikivski2004}. Self-avoiding (SA) polymer models have been studied numerically and compared with the flexible and semiflexible ideal chain models in Refs. \ref{Shin2010} and \ref{Mokkonen2014}. SA polymers show qualitatively similar behaviour as ideal polymers, but unlike the monotonically decreasing escape rate of ideal polymers, the escape rate of the SA polymers exhibits a minimum for intermediate length beyond which the escape rate increases. 

In this article, we study polymer escape dynamics and estimate the escape rate using harmonic approximation of the transition state theory followed by dynamical corrections (DC) for a discrete, one-dimensional harmonic ideal polymer model in a symmetric double-well external potential. We compare the results with molecular dynamics simulations using both Brownian dynamics (BD) and Langevin dynamics (LD). We also compare with results obtained from Langer's rate estimate \cite{Lee2001, Sebastian2006}. Using the Nudged Elastic Band (NEB) method, we find the minimum energy paths (MEP) to identify the relevant saddle points on the energy surface as a function of chain length $N$ and polymer spring constant $K$. The harmonic modes at the saddle point are evaluated and their eigenvalues and eigenvectors used to analyse the escape dynamics of the polymer.

The system studied is described in Sec. \ref{sec:system}, the methods for calculating the rate in Sec. \ref{sec:rate} and the numerical methods in Sec. \ref{sec:numerical}. The results are presented in Sec. \ref{sec:results}. The article concludes with a summary and discussion in Sec. \ref{sec:discussion}.

%%%%%%%%%%%%%%%%%%%%
%  SYSTEM					  %
%%%%%%%%%%%%%%%%%%%%

\section{System Description and Problem Definition} \label{sec:system}

We consider a polymer represented by $N$ beads in one dimension interacting with their neighbours through a spring.  The configuration of the polymer is described by the set of coordinates $\mathbf{r} :=  \{ r_n \}_{n=1}^N$ with the centre of mass $\frac{1}{N} \sum_{n=1}^N r_n$. The equation of motion for the $n$th bead at time $t$ is given by Langevin dynamics (LD) as
\begin{equation} 
m \ddot{r}_n(t) +  \gamma \dot r _n(t)  + \nabla_n [V(r_n(t)) + U] = \sqrt{2 \gamma k_B T } \xi_n (t), \label{eq:langevin}
\end{equation} 
where $m$ is the mass of an individual bead, $\gamma$ the friction coefficient, $V(
r_n)$ the external potential, $U$ the interaction potential between beads, $k_BT$ the thermal energy, and $\xi_n (t)$ a Gaussian random force satisfying $\langle \xi_n (t) \rangle = 0$, and $\langle \xi_n (t) \xi_m (t') \rangle = \delta (t - t') \delta_{n,m}$. In the limit of strong coupling with the heat bath, i.e. in the high friction limit, the dynamics become overdamped and are described by the Brownian equation of motion (BD)
\begin{equation} 
\gamma \dot r _n(t)  + \nabla_n [V(r_n(t)) + U] = \sqrt{2 \gamma k_B T } \xi_n (t). \label{eq:brownian}
\end{equation} 

\begin{figure}
\centering
\includegraphics[width=171mm]{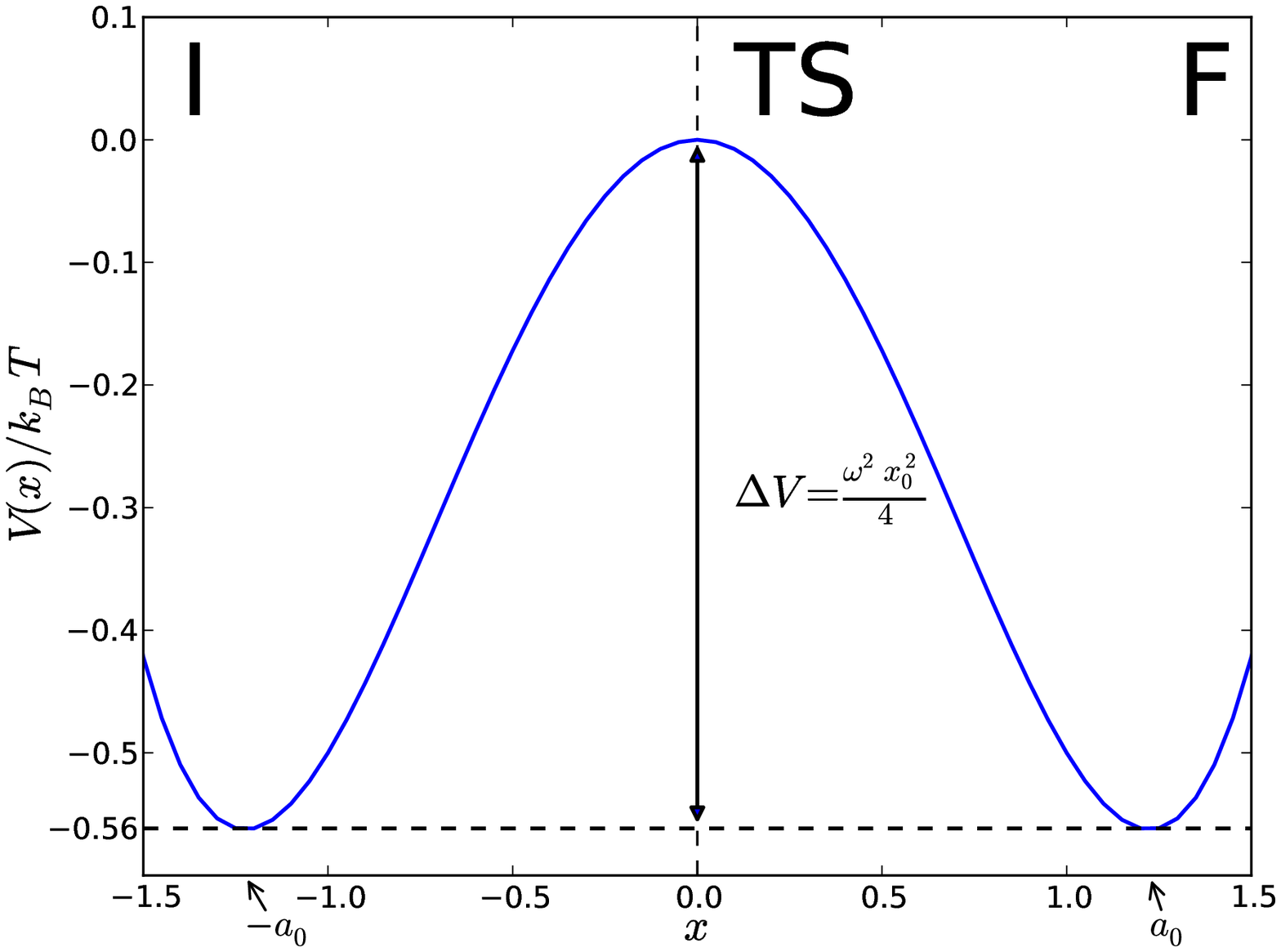}
\caption{The external potential of Eq. \eqref{eq:externalpotential}. The maximum of height $\Delta V = \omega^2 a_0^2 / 4 \approx 0.56$ is located at $x = 0$ and the minima are located at $x = \pm a_0 \approx \pm 1.22$. The  initial state, I, is confined to the left well $x < 0$ and the right well $x > 0$ corresponds to the final state, F. In the multidimensional transition state for a polymer escape event, the center of mass of the polymer is located at the maximum of the external potential, as indicated by the label TS.} \label{fig:externalpotential}
\end{figure}

The interaction potential between the  beads is given by a harmonic potential $U =  \sum_{n=1}^{N-1}(K/2)(r_n - r_{n+1})^2$ giving the force 
\begin{equation} 
- \nabla_n U = - K (r_{n-1} + r_{n+1} - 2r_n),
\end{equation}
acting on the $n$th bead. The external potential $V(x)$, shown in Fig. \ref{fig:externalpotential}, is a quartic double well
\begin{equation} 
V(x) = - \frac{\omega^2}{2} x^2 + \frac{\omega^2}{4 a_0^2} x^4,\label{eq:externalpotential}
\end{equation}
where $\pm a_0$ gives the location of the minima, the energy has a maximum at $x=0$, and $\omega^2$ is the curvature of the energy barrier. The same polymer model and external potential with BD were used in Ref. \ref{Lee2001}. The energy of a polymer configuration $\mathbf{r}$ is given by 
\begin{equation} 
\Phi (\mathbf r) = \sum_{n=1}^N V(r_n)  +  \sum_{n=1}^{N-1}(K/2)(r_n - r_{n+1})^2. \label{eq:harmonicforce}
\end{equation}

In the one-dimensional Rouse model, a continuum model of an ideal chain, the spring constant is $K = k_B T /  l_0^2$ which gives the polymer an average bond length $l_0 = \sqrt{k_B T / K}$ and an average squared end-to-end distance of $\langle R_\mathrm{ee}^2 \rangle =  N l_0^2$ in free space. 
The parameter $\bar R_\mathrm{ee} = \sqrt{\langle R_\mathrm{ee}^2 \rangle}$ can be used to compare the size of the polymer to the width of the potential well.
\cite{Doi1986}

At time $t = 0$, all the beads of the polymer are in the left potential well shown in Fig. \ref{fig:externalpotential}, so $X_0 < 0$. The polymer is thermally equilibrated with the heat bath and its configuration consistent with Boltzmann distribution $P_\mathrm{I} (\mathbf r) = P(\mathbf r | X_0 < 0) \propto \exp(- \Phi (\mathbf r) / k_B T)$. With time evolution according to either Eq. \eqref{eq:langevin} or Eq. \eqref{eq:brownian}, the polymer eventually escapes to the final state F, where $X_0 > 0$. The rate of such escape events is $\mathcal R_{\mathrm I \rightarrow \mathrm F}$ and the time in between escape events is $t_\mathrm{cross} \approx \mathcal R^{-1}_{\mathrm I \rightarrow \mathrm F} $ on average. 

%%%%%%%%%%%%%%%%%%%%
%  CALCULATION OF THE RATE	   %
%%%%%%%%%%%%%%%%%%%%

\section{Calculation of the Rate} \label{sec:rate}

\subsection{Molecular Dynamics Simulations}

The LD Eq. \eqref{eq:langevin} or the BD Eq. \eqref{eq:brownian} are integrated numerically to generate trajectories representing transitions from I to F. Uncorrelated samples of the Boltzmann distribution $P_\mathrm{I} (\mathbf r)$ for the polymer in state I were generated by simulating the chain confined to the left potential well for time intervals twice as long as the correlation time between samples. From these trajectories, the escape probability was determined as
\begin{equation}
P_\mathrm{esc} (t) = (1/N_\mathrm{traj}) \sum_{i=1}^{N_\mathrm{traj}} \theta(t - t_i), \label{eq:pesc}
\end{equation} 
where $N_\mathrm{traj}$ is the number of simulated trajectories, $\theta(t -t_i)$ the Heaviside step function and $t_i$ the time of the $i$th escape event. {An escape event is considered to have occured when the centre of mass of the system $X_0$ has reached $X_0 > a_0/2$.} The dynamical escape rate is given by 
\begin{equation}
\mathcal R_\mathrm{MD} = \left. \frac{d P_\mathrm{esc} (t)}{dt} \right., \label{eq:dynamicalrate}
\end{equation} 
where the derivative is computed by fitting the curve $P_\mathrm{esc} = \mathcal R_\mathrm{MD} t + \mathrm{b}$, where $\mathrm{b}$ is a constant, over the time interval where $P_\mathrm{esc} (t)$ is close to being a linear function of $t$. The average time in between escape events is $t_\mathrm{cross}  = \langle t_i \rangle = (1/N_\mathrm{traj}) \sum_{i=1}^{N_\mathrm{traj}}  t_i \label{eq:tcross}$.

\subsection{Transition State Theory}

Transition state theory (TST) assumes that the initial distribution has the Boltzmann form $P_\mathrm{I} (\mathbf r) \propto \exp(-\Phi (\mathbf r) / k_B T)$ and that the energy barrier is high enough for the time between escape events to be longer than the relaxation time. A transition state is defined in between the initial and final states that should represent a bottleneck for the transition. The key assumption of TST is the no-recrossing approximation where a trajectory crossing the transition state with velocity pointing away from the initial state is assumed to end up in the final state without recrossings. Implicit in this approximation is an assumption that the time spent in the vicinity of the transition state is short compared to the time spent in the final state.  The TST rate estimate is obtained by multiplying the probability that the system reaches the TS with the flux through the TS 
\begin{equation}
\mathcal{R}_\mathrm{TST} = \sqrt{k_B T / (2 \pi \mu_\perp)} Z_\mathrm{TS}/Z_\mathrm{I}, 
\end{equation}
where $\mu_\perp$ is the reduced mass of the system in the direction perpendicular to the TS surface, and $Z_\mathrm{TS}$ and $Z_\mathrm{I}$ are the configuration integrals of the TS and I, respectively. \cite{Wigner1935}

Instead of computing the full configuration integrals $Z_\mathrm{TS}$ and $Z_\mathrm{I}$, a harmonic approximation to TST (HTST) can be used \cite{Wert1949}. If the escape trajectories are likely to go through the vicinity of a first order saddle point on the energy surface in between the energy minima corresponding to the I and F states, the configuration integrals can be evaluated analytically by making a second order expansion around the saddle point and the initial state minimum. A convenient choice for a reaction coordinate is the minimum energy path (MEP) and the relevant saddle point is the point of maximum energy along the MEP. In HTST, the TS is chosen to be a hyperplane going through the saddle point with its normal pointing along the MEP. The MEP starts from $\mathbf r _ \mathrm I  = - a_0 \mathbf 1 $, where $\mathbf 1 = [1,1, \dots, 1]$ is a vector of length $N$, and ends at $\mathbf r _ \mathrm F  = a_0 \mathbf 1$ and the location of the saddle point is dennoted by $\mathbf r_\ddagger$. After expanding the total energy of the system up to second order around the initial state minimum $\mathbf{r}_0$ and the saddle point $\mathbf{r}_\ddagger$, the configuration integrals $Z_\mathrm{TS}$ and $Z_\mathrm{I}$ can be evaluated analytically to give the HTST estimate of the transition rate
\begin{equation}
\mathcal{R}_\text{HTST} = \frac{1}{2 \pi \sqrt{\mu_\perp}} \sqrt{\frac{\prod_{i=1}^N \lambda^0_i}{\prod_{i=2}^{N} \lambda^\ddagger_i} }e^{-\Delta \Phi / k_B T}. \label{eq:ratehtst}
\end{equation}
Here $ \lambda_i^0$ and $ \lambda^\ddagger_i$ are the eigenvalues of Hessian matrix which has elements
\begin{equation}
(H_p)_{ij} = \left. \frac{\partial^2 \Phi (\mathbf r)}{\partial r_i \partial r_j} \right|_{\mathbf r = \mathbf r _p}, \label{eq:hessian}
\end{equation}
evaluated at the initial state minimum and at the saddle point. The product in the denominator of the HTST rate expression omits the negative eigenvalue corresponding to the direction along the MEP. The elements of the Hessian can be computed using finite differences of the forces acting on the beads. Diagonalisation of the Hessian matrix of Eq. \eqref{eq:hessian} gives an eigenvalue spectrum $\text{diag}(\lambda_i) = \mathbf{S}^{-1} \mathbf{H} \mathbf{S}$ and a set of eigenvectors, $\mathbf{S} = [\mathbf s_1 \dots \mathbf s_N]$, which are referred to as the harmonic modes.

The configuration of the polymer at the saddle point depends on the chain length in such a way that below a crossover length $N_C$ the polymer is collapsed to one point $\mathbf{r}_\ddagger = \mathbf{0}$. For longer polymers with $N>N_C$, it becomes energetically more favourable for the chain to be stretched along the MEP. This transition from collapsed to stretched polymer is analogous to the onset of quantum mechanical tunneling and the appearance of an instanton in quantum rate theories based on Feynman path integrals. The relation between the barrier curvature, spring constant and critical chain length can be derived analytically from the continuum model of an ideal chain \cite{Lee2001}:
\begin{equation}
N_C = \sqrt{\frac{K \pi^2}{\omega^2}}. \label{eq:nc}
\end{equation}

At $N_C$, the eigenvalue of the smallest positive mode at the saddle point, $\lambda^\ddagger_2$, approaches zero. This leads to a divergence in the HTST rate estimate which can be reduced by introducing anharmonic corrections (AHC) for this mode. Lee and Sung have derived an expression for the correction factor as
\begin{equation}
g(\alpha) = \sqrt{\frac{\alpha}{2 \pi}} \int_{-\infty}^{\infty} dQ e^{-(\alpha/2)Q^2 - (3/8) Q^4}, \label{eq:anharmcorr}
\end{equation}
where $\alpha =  \lambda^\ddagger_2 a_0  \sqrt{N/(k_B T)}/ \omega$ \cite{Lee2001}. Thus, the corrected HTST rate becomes
\begin{equation}
\mathcal{R}_\text{HTST+AHC} = g(\alpha) \mathcal{R}_\text{HTST}.  \label{eq:ratehtstanharm}
\end{equation}

\subsection{Dynamical Corrections}

The assumption of no recrossings of the TS in TST is often not satisfied for energy barriers that are broad. The effect of the recrossings can, however, be estimated using calculations of classical trajectories started at the TS. {Initial configurations of the trajectories are sampled within the TS hyperplane, such that net force is zero $\mathbf F_\mathrm{0} =  \mathbf F -  \sum_{i=0}^N F_i/N = 0$ which keeps the system within the plane.} This provides dynamical corrections (DC) to the TST rate estimate. The correction factor, $\kappa$, is \cite{Voter1985}:
\begin{equation}
\kappa = \frac{2}{N_\mathrm{traj}}\sum_{i=1}^{N_\mathrm{traj}} \mathrm{sgn}(v_i) \theta(X_0^\mathrm{final}),
\end{equation}
where $\mathrm{sgn}(v_i)$ is the sign of the initial net velocity $v_i = \sum_{j=0}^N v_j$ of a trajectory $i$ assigned from the Maxwell-Boltzmann distribution $P(v) \propto \exp [ -v^2 / (2 k_B T)]$, and $\theta(X_0^\mathrm{final})$ is unity if the system resides in F at the end of the trajectory and zero otherwise. Thus, each trajectory that ends in F, contributes positively to $\kappa$ if it starts with a velocity pointing towards F, but contributes negatively if the initial velocity points towards I. Trajectories ending in I do not contribute to $\kappa$. The corrected rate estimate is
\begin{equation}
\mathcal{R}_\text{HTST+AHC+DC} = \kappa \mathcal{R}_\text{HTST+AHC},  \label{eq:ratedc}
\end{equation}
where $\mathcal{R}_\text{HTST+AHC}$ is given by Eq. \eqref{eq:ratehtstanharm}.
While the direct dynamical calculations of transitions can be impossibly long for systems with low transition rates, the calculation of DC requires only short trajectories and can be carried out readily with little computational effort.

\subsection{Kramers' and Langer's Rate Theory}

Langer's rate theory for barrier crossing is a multidimensional generalisation of Kramers' canonical rate theory \cite{Langer1969}. Langer's expression for the rate resembles the HTST rate expression, although it is derived from different assumptions. The main difference between Langers' and HTST rate estimates is that Langer's rate has an additional pre-exponential factor that provides an approximate estimate of the effect of TS recrossings. Langer's rate estimate is:
\begin{equation}
\mathcal{R}_\text{L} = {\frac{ \sqrt{|\lambda^\ddagger_1|} }{2 \pi \gamma}} \sqrt{\frac{\prod_{i=1}^N \lambda^0_i}{\prod_{i=2}^{N} \lambda^\ddagger_i}}  e^{-\Delta \Phi / k_B T}. \label{eq:ratelanger} 
\end{equation}
Langer's approach was used by Sebastian \textit{et al.} and Lee and Sung \cite{Sebastian2000a, Sebastian2006,Lee2001}. They calculated the unstable mode and partition functions analytically for the Rouse model, a continuum model of an ideal chain. Langer's rate estimate diverges around $N_C$ similar to the HTST rate estimate and this divergence can be removed using the correction factor of Eq. \eqref{eq:ratehtstanharm}.

%%%%%%%%%%%%%%%%%%%%
%  SIMULATIONS				  %
%%%%%%%%%%%%%%%%%%%%

\section{Numerical Methods} \label{sec:numerical}

The escape rate was obtained by numerical molecular dynamics simulations using both Langevin dynamics (LD) of Eq. \eqref{eq:langevin} and Brownian dynamics (BD) of Eq. \eqref{eq:brownian}. BBK integration \cite{Brunger1984} was used for LD  with a time step of $\Delta t = 0.005$, and forward Euler integration for BD with time step ranging between $0.005$ and $0.01$. Chains with different combinations of $N$ and $K$ in ranges $N \in \{1, \dots, 120\}$ and $K \in \{5, \dots, 60\}$ were simulated with parameters $\gamma = 1.0$, $m=1.0$ and $k_B T = 1.0$. The parameters for the external potential of Eq. \eqref{eq:externalpotential} were $\omega^2 = 1.5$ and $a_0^2 = 1.5$. The parameters are the same as those used in Ref. \ref{Lee2001}. If we choose the units of length, mass and energy to be $l_0 = 1.02$ nm, $m_0 = 1870$ amu, corresponding to double stranded DNA, and $k_B T$ at $T = 300$ K, the unit of time becomes $t_0 = \sqrt{m_0 l_0^2 / k_B T}  =27.9$ ps.

In order to find the relevant saddle point to apply HTST of Eq. \eqref{eq:ratehtst} the Nudged Elastic Band (NEB) method \cite{Jonsson1998, Henkelman2000} was used to find the MEP of the escape transition. In the NEB method, a set of replicas of the system, referred to as images, $\{ \mathbf r_p \}_{p=1}^P$ are placed along a path between the initial state minimum $\mathbf r _ \mathrm I  = - a_0 \mathbf 1$ and the final state minimum $\mathbf r _ \mathrm F  =  a_0 \mathbf 1$. The images represent a discretisation of the path and to control the distribution of these discretisation points, the images are connected with harmonic springs with spring constant $k_\mathrm{NEB}$. An estimate of the tangent \cite{Henkelman2000} to the path is used to project out the parallel component of the true force acting on the beads and the perpendicular component of the spring force (the 'nudging'). The images are then displaced iteratively so as to zero the net force acting on the beads until the images lie along the MEP. 

At the end of the NEB calculation the images give a discrete representation of the MEP, but no image will be exactly at the saddle point. An accurate estimate of the saddle point was found here by identifying the image with the highest energy and then minimising the force on this image with the Newton-Raphson method. The method consists of iterations $\mathbf{r}^{(n+1)} = \mathbf{r}^{(n)} - \mathbf{H}^{-1} \mathbf{F}^\mathrm{sys} (\mathbf{r}^{(n)})$, where $\mathbf{F}^\mathrm{sys} (\mathbf{r})$ is the force of the real system (no NEB spring forces) and {$\mathbf{H}^{-1}$ is an inverse of the Hessian matrix (Jacobian matrix of the force) with elements given by Eq. \eqref{eq:hessian}.} %$\text{J}_{ij} = \partial \mathbf{F}^\mathrm{sys}(r_i) /\partial r_j$. 
Within a few iterations, the method converges to a point $\mathbf{r}_\ddagger$ where $\mathbf{F} (\mathbf{r}_\ddagger) =  0$. If the initial point is close enough to a saddle point, then the Newton-Raphson method will most likely converge onto the saddle point rather than other points where the force is zero. The activation energy is then given by the energy difference between the initial state and the saddle point as $\Delta \Phi = \Phi(\mathbf r_\ddagger) - \Phi(\mathbf r_0)$. The Hessian matrices $\mathbf H_0$ and $\mathbf H_\ddagger$ of Eq. \eqref{eq:hessian} can be evaluated using the finite difference method. {Alternatively, the exact saddle point can be found with the climbing image method \cite{Henkelman2000a}.}

The spring constants used in the NEB calculations was $k_\mathrm{NEB} = 8.2 \times 10^{-5}$ for $K=10$, and $k_\mathrm{NEB} = 8.2$ for $K=60$. The number of images, $P$, was typically chosen to be between 9 and 19, but for the larger $N$ sometimes up to $P = N/2$.

%%%%%%%%%%%%%%%%%%%%
%  RESULTS		           		  %
%%%%%%%%%%%%%%%%%%%%
 
\section{Results} \label{sec:results}

\subsection{Minimum Energy Paths and Saddle Points}

\begin{figure}
\centering
\includegraphics[width=120mm]{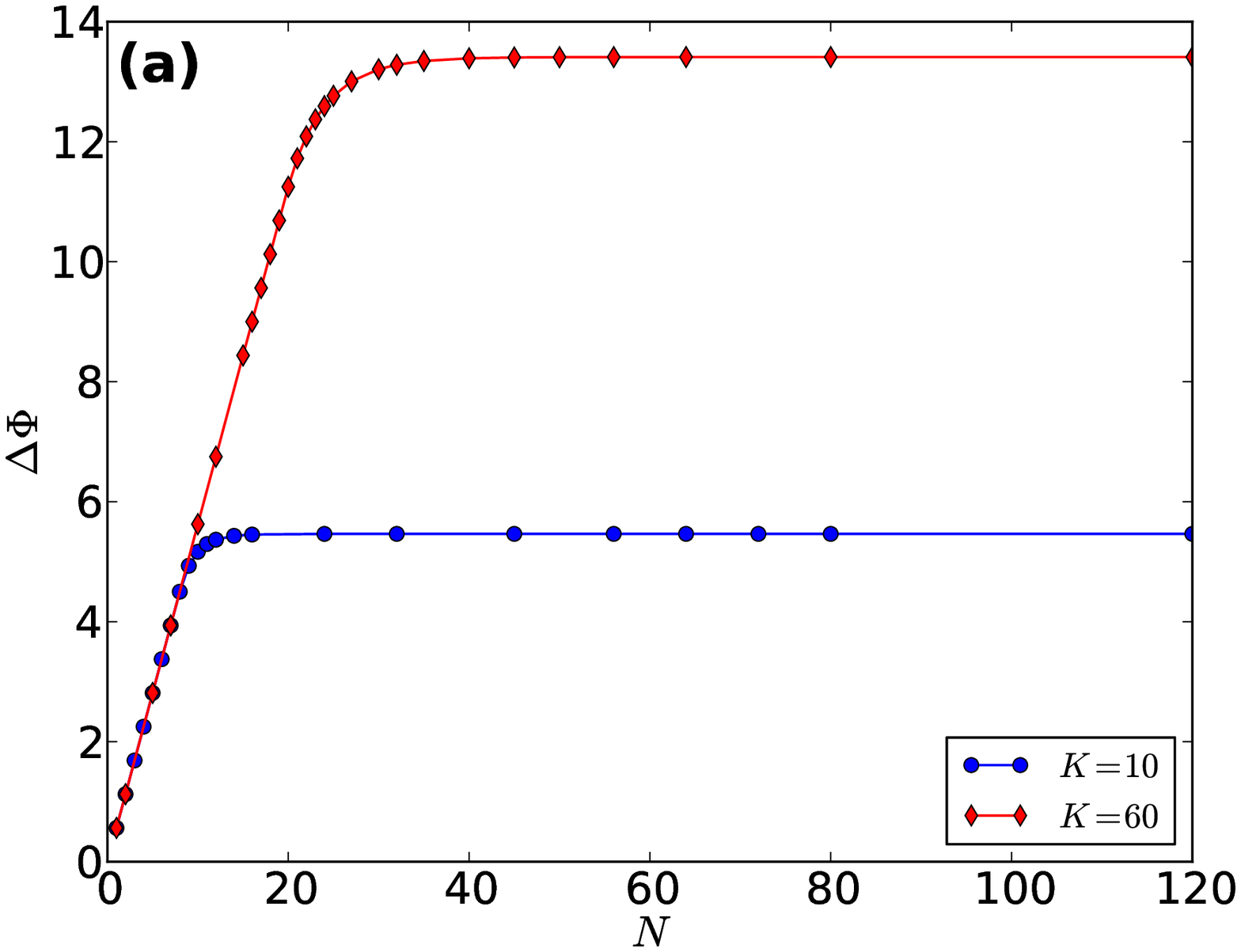} 
\includegraphics[width=120mm]{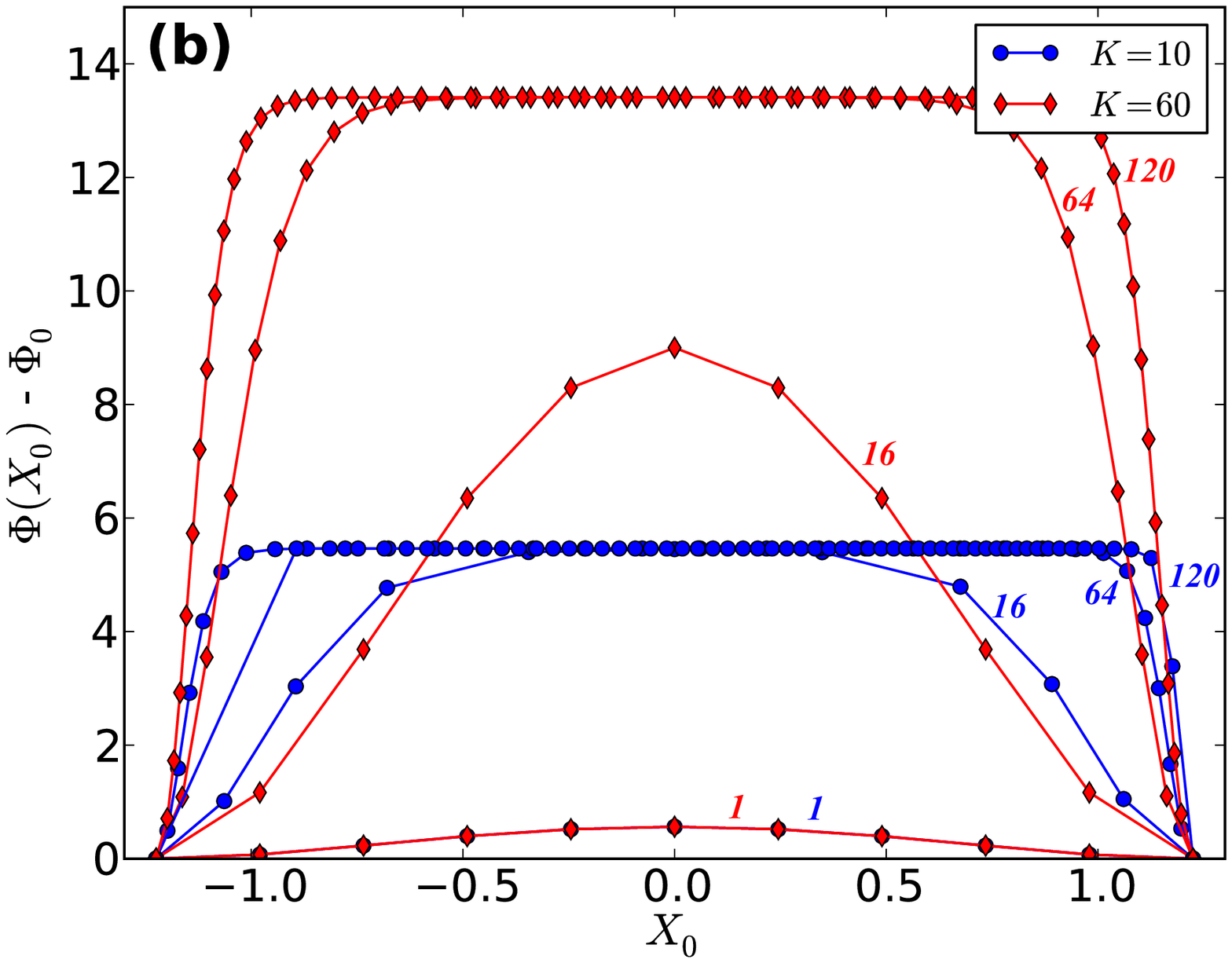} 
\caption{(a) The activation energy given by the maximum energy along the minimum energy path on the energy surface $\Delta \Phi $ as a function of $N$ for $K=10$ (circles) and for $K=60$ (diamonds). The activation energy  increases linearly with $N$ until $\tilde N_C$, the onset of stretched polymer configurations, beyond which it remains constant.  (b) Energy along MEPs vs. location of the center of mass of the polymer, $X_0$, for $N=\{1,16,64,120\}$ $K=10$ (circles) and for $K=60$ (diamonds). The energy at the flat region becomes independent of $X_0$ over a range of values for $N>{\tilde N_C}$}\label{fig:meps}
\end{figure}

The energy along minimum energy paths for polymer escape are shown in Fig. \ref{fig:meps}.a. The energy barrier is found to increase linearly with $N$ up to $\tilde N_C$ (the integer value of the crossover length obtained from MEP calculations is denoted by $\tilde N_C$ while the estimate obtained from the continuous ideal chain, Eq. \eqref{eq:nc}, is denoted by $N_C$) where it reaches a constant value. Below $\tilde N_C$ the saddle point configuration corresponds to a coiled up polymer where all the beads lie on top of the maximum of the external potential at $x=0$, and the height of the energy barrier is $\Delta \Phi = N \Delta V$. Beyond $\tilde N_C$, the stretched configuration is energetically more favourable. The values of $\tilde N_C$ obtained from MEPs are $\tilde N_C = 8$ for $K=10$ and $\tilde N_C = 20$ for $K=10$, which agree well with theoretical predictions of Eq. \eqref{eq:nc}: $N_C = 8.11$ for $K=10$ and $N_C = 19.86$ for $K=60$. 

\begin{figure}
\centering
\includegraphics[width=110mm]{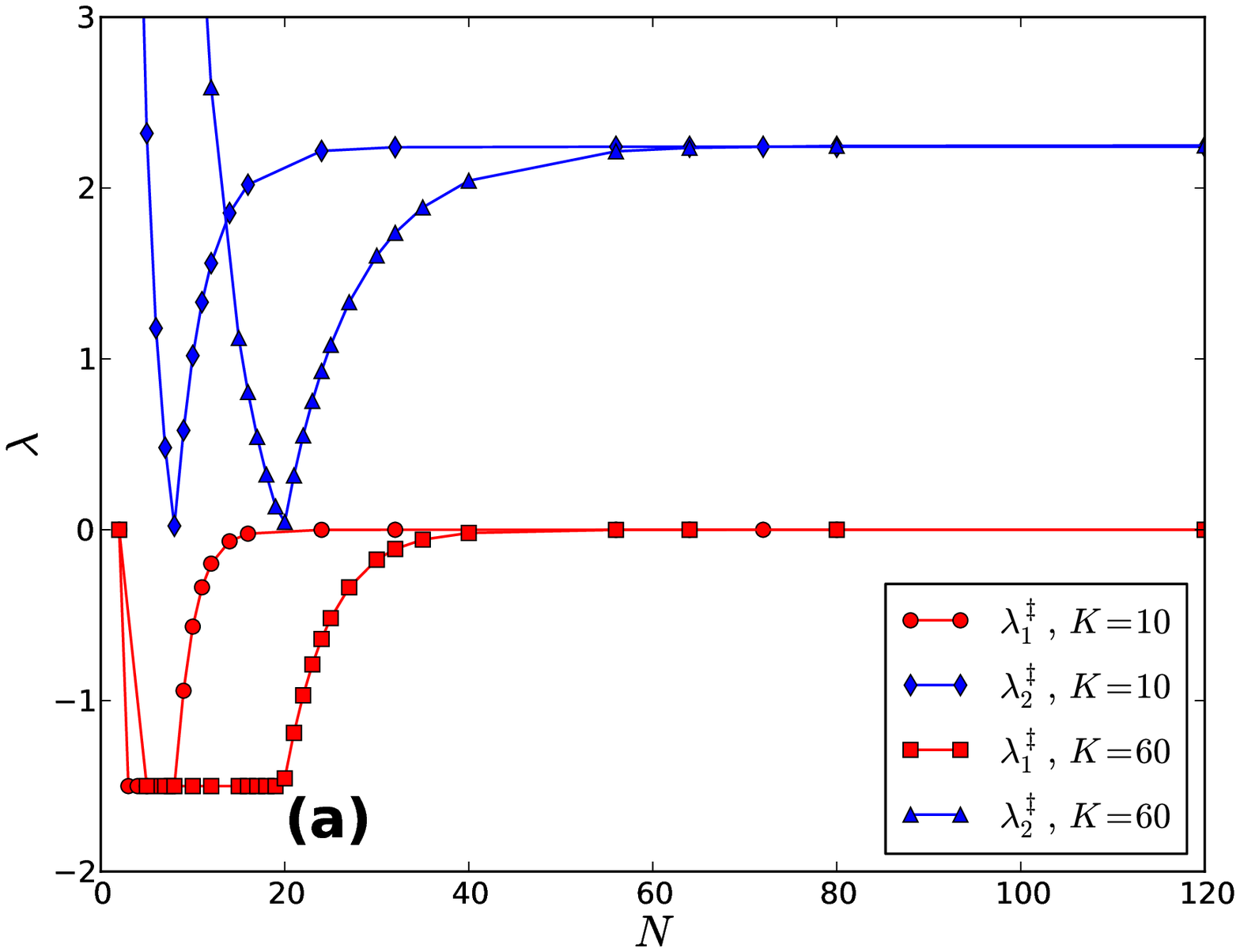}   
\includegraphics[width=110mm]{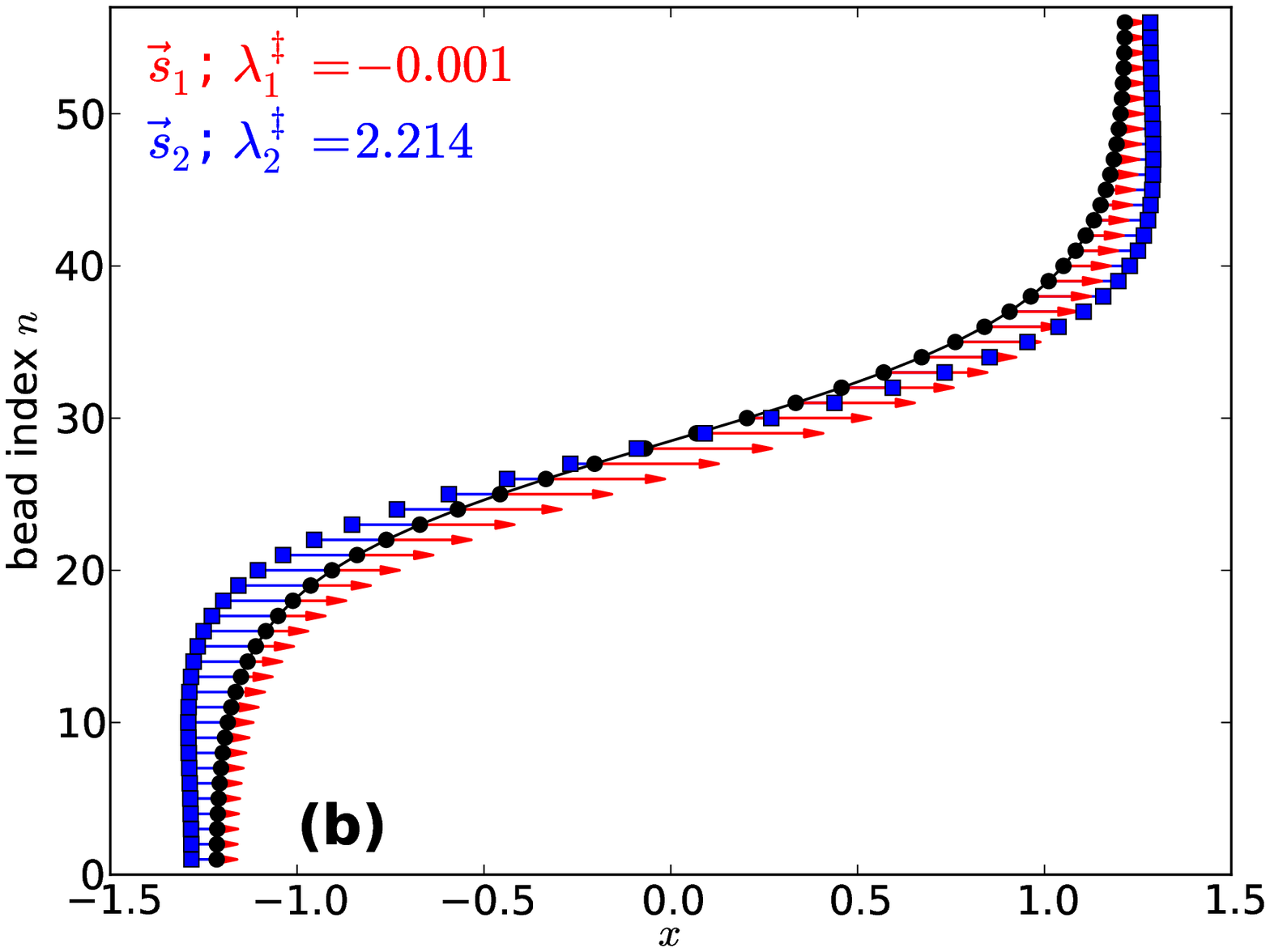} 
\caption{(a) The two lowest eigenvalues of the Hessian, Eq. \eqref{eq:hessian}, as a function of chain length $N$. The negative eigenvalue $\lambda^\ddagger_1$ corresponding to the unstable mode approaches zero as $N$ increases beyond $\tilde N_C$. The smallest positive eigenvalue, $\lambda^\ddagger_2$, has a minimum of near zero at $\tilde N_C$ and then reaches a constant value as $N$ increases beyond $\tilde N_C$. (b) A saddle point configuration and the modes corresponding to the two lowest eigenvalues of a polymer with $N=56$ beads and spring constant $K=60$. The location of the beads is marked with black dots and the normalised eigenvectors indicated, $\mathbf s_1$ with triangle capped arrows and $\mathbf s_2 $ with square capped arrows. The $\mathbf s_1$ mode corresponds to motion along the minimum energy path towards the final state and the mode $\mathbf s_2 $ corresponds to the stretching of the polymer. } \label{fig:modes}
\end{figure}

The eigenvalue of the smallest positive mode at the saddle point, $\lambda_2^\ddagger$, decreases with $N$ until $ \tilde N_C$, where it has a minimum of nearly zero.  Beyond $ \tilde N_C$ it increases, as shown in Fig. \ref{fig:modes}.a.  The mode corresponding to this eigenvalue is responsible for creating the stretched configuration; instead of the eigenvalue $\lambda_2^\ddagger$ becoming negative and the mode unstable, the polymer stretches along the energy barrier of the external potential, extending the springs between beads near the barrier top but at the same time lowering the number of beads at the barrier top and thereby reducing the overall activation energy below $N \Delta V$. The stretched configuration for the chain of length $N=56$ is shown in Fig. \ref{fig:modes}.b along with the modes corresponding the eigenvalues $\lambda_1^\ddagger$ and $\lambda_2^\ddagger$.

The mode with the negative eigenvalue $\lambda_1^\ddagger$ corresponds to the polymer moving towards the final state as shown in Fig. \ref{fig:modes}.b. Below $\tilde N_C$, the negative eigenvalue $\lambda_1^\ddagger$ corresponds to the curvature of the external potential $\omega^2$. At $\tilde N_C$, the negative eigenvalue $\lambda_1^\ddagger$  starts increasing and approaches zero with increasing $N$, because the energy along the MEP flattens out. Fig. \ref{fig:meps}.b shows the energy along the MEP for various values of $N$ illustrating how the curvature at the saddle point decreases with increasing chain length.

\subsection{Brownian and Langevin Dynamics}

The molecular dynamics simulations were performed using both Langevin dynamics of Eq. \eqref{eq:langevin} and Brownian dynamics of Eq. \eqref{eq:brownian}. An escape event was considered to have occured when the centre of mass of the system $X_0$ had reached {$X_0=a_0/2 \approx 0.61$. As can be seen in Fig. \ref{fig:meps}.b, the barrier is flat in this region for the longer chains. We repeated simulations for chains $N \in \{56,64,80\}$ with $K=10$ and the crossing condition $X_0=1.0$, and found the difference in the calculated rates to be negligible.}

\begin{figure}
\centering
\includegraphics[width=171mm]{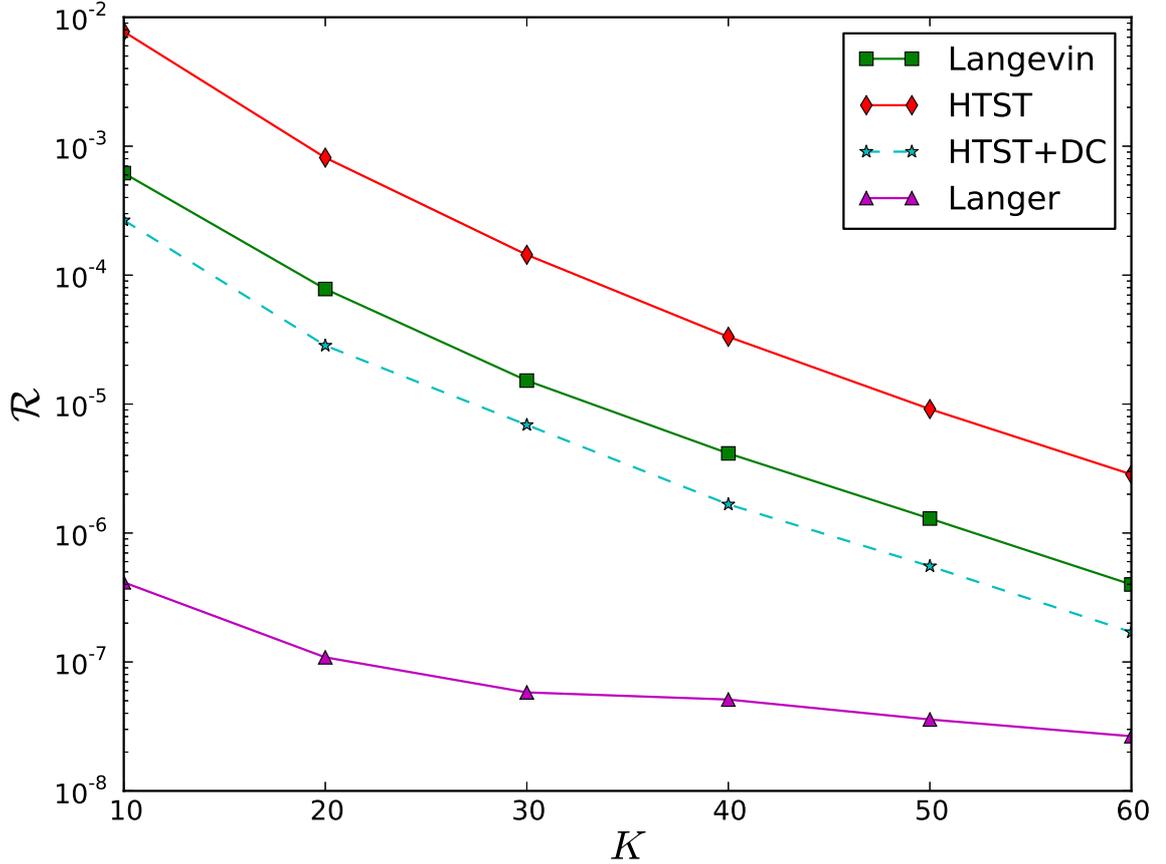} 

\caption{The rate of escape of a polymer of length $N=64$ as a function of spring constant, $K$. Squares indicate the rates obtained from Langevin dynamics simulations, circles the HTST estimates, Eq. \eqref{eq:ratehtst}, stars the  estimates from HTST with dynamical corrections, HTST+DC from \eqref{eq:ratedc}, triangles the estimates from Langer's theory, Eq. \eqref{eq:ratelanger}. The HTST+DC rate estimate is in good agreement with the rate obtained from the dynamics simulations, but Langer's theory significantly underestimates the rate because the eigenvalue corresponding to the unstable mode approaches zero for long polymers.
} 
\label{fig:ratesK}
\end{figure}

The escape rate obtained from these simulations using Eq. \eqref{eq:pesc} is shown in Fig. \ref{fig:rates} as a function of $N$ for chains with $K=10$ and $K=60$ and as a function of the spring constant $K$ for chain of length $N=64$ in Fig. \ref{fig:ratesK}. The maximum chain length for for which LD simulations were carried out was limited by computational resources. The escape trajectories were started from the equilibrated initial distribution $P_\mathrm{I} (\mathbf r) \propto \exp(- \Phi (\mathbf r) / k_B T)$ with the constraint $X_0 < 0$ for both LD and MD. Lee and Sung used the initial condition $\mathbf r = - a_0 \mathbf 1$ \cite{Lee2001} instead of trajectories started from a Boltzmann distribution. We compared these two initial conditions and found no difference in the escape rate.

Both escape rates show a monotonically decreasing rate in both $N$ and $K$, which is consistent with previously published work \cite{Lee2001, Shin2010, Mokkonen2014}. Our estimates of the escape rate obtained from BD agree well with the results reported by of Lee and Sung \cite{Lee2001}. The rate obtained from BD trajectories is slightly higher than the rate obtained from LD trajectories. 
{We note that the BD and LD rates are not the same because of the lack of inertial effects in BD. Nevertheless, with the current
parameters the LD simulation results are very close to the overdamped BD case. }

\subsection{Transition State Theory with Dynamical Corrections}

\begin{figure}
\centering
\includegraphics[width=90mm]{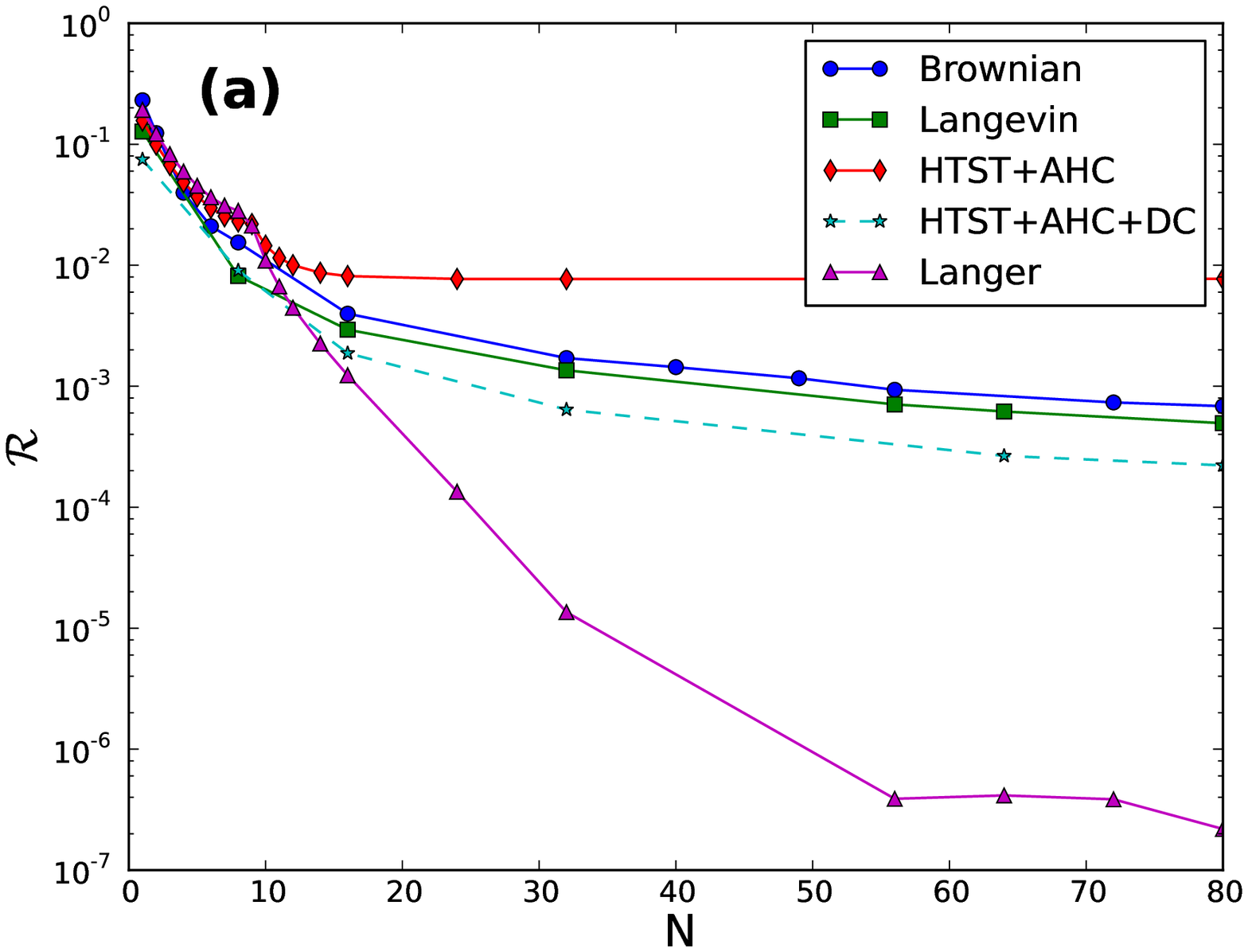}
\includegraphics[width=90mm]{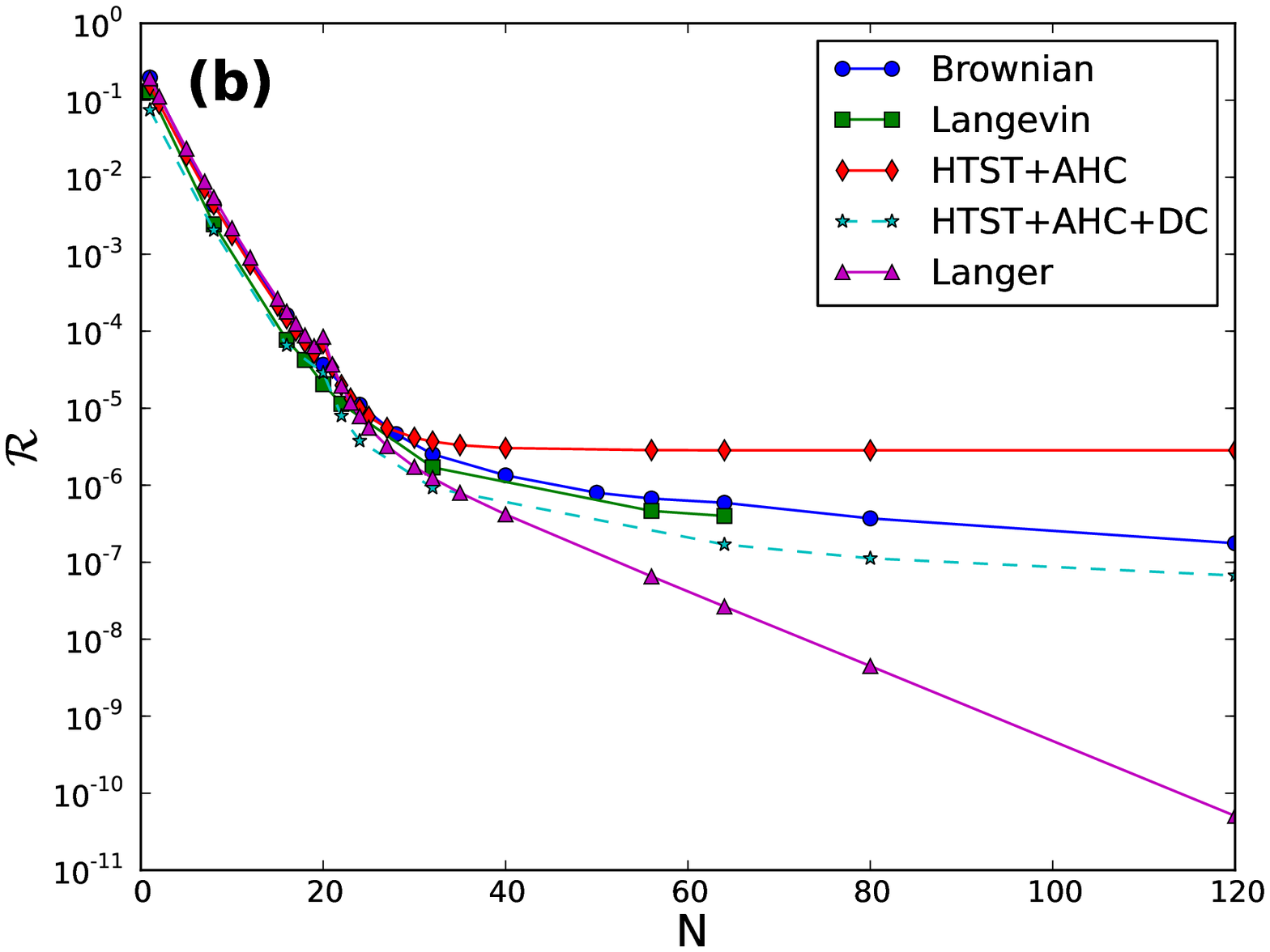}  
\caption{Rate of polymer escape as a function of chain length $N$ for two values of the spring constant: (a) $K=10$ and (b) $K=60$. Circles and squares are results of direct simulations using Brownian and Langevin dynamics, respectively. Diamonds indicate rate estimates obtained from harmonic transition state theory with anharmonic corrections, HTST+AHC, for polymers with length close to $\tilde N_C$ to remove divergence caused by the zero in the smallest positive eigenvalue of the Hessian at the saddle point. Stars indicate rate estimates obtained after correcting for recrossings, HTST+AHC+DC, using short time trajectories started at the transition state. The rate estimates obtained from Langer's theory are indicated with triangles, showing a large underestimate of the rate for long chains. The eigenvalue $\lambda^\ddagger_1$ is less than $10^{-9}$ for $N>56$ and $K=10$ which causes problematic convergence of NEB along this mode leading to numerical inaccuracies in Langer's rate.
} \label{fig:rates}
\end{figure}

\begin{figure}
\centering
\includegraphics[width=171mm]{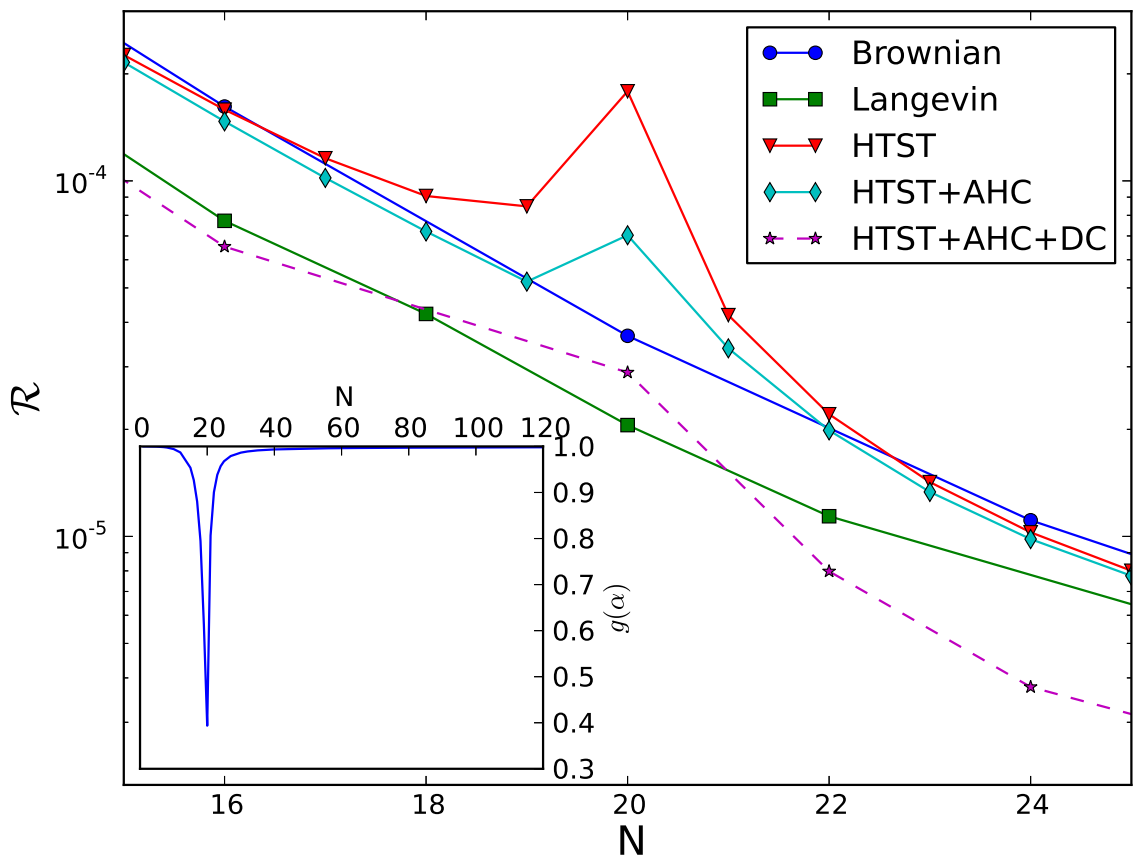} 
\caption{The escape rate calculated for polymers with spring constant of $K=60$ as a function of length close to the crossover, $\tilde N_C$. Circles and squares indicate results obtained from dynamical simulations using Brownian and Langevin dynamics, respectively. Triangles indicate the HTST rate estimates, Eq. \eqref{eq:ratehtst}, diamonds the estimates from HTST with anharmonic corrections, Eq. \eqref{eq:ratehtstanharm}, and stars the estimate with dynamical corrections, Eq. \eqref{eq:ratedc}. The anharmonic correction factor, $g(\alpha)$ given by Eq.  \eqref{eq:anharmcorr}, shown in the inset, significantly reduces the diverging peak in the HTST rate estimate.} 
\label{fig:ratesK60_zoom}
\end{figure}

The escape rate $\mathcal R_\mathrm{HTST+AHC}$ obtained by harmonic transition state theory (HTST) with anharmonic corrections (AHC) of Eq. \eqref{eq:ratehtstanharm}, and $\mathcal R_\mathrm{HTST+AHC+DC}$ with dynamical corrections (DC) of Eq. \eqref{eq:ratedc}, are shown in Fig. \ref{fig:rates} as a function of $N$ for chains with spring constants $K=10$ and $K=60$. In Fig. \ref{fig:ratesK}, $\mathcal R_\mathrm{HTST+AHC}$ and $\mathcal R_\mathrm{HTST+AHC+DC}$ are shown as a function of the spring constant $K$ for a chain of length $N=64$. The peak at $\tilde N_C$ arising from Eq. \eqref{eq:ratehtst} is removed from the rate in Fig. \ref{fig:rates} using the anharmonic correction of Eqs. \eqref{eq:anharmcorr} and \eqref{eq:ratehtstanharm}. The HTST rate of Eq. \eqref{eq:ratehtst} without AHC is compared with the corrected rate in Fig. \ref{fig:ratesK60_zoom}. The peak arises in the HTST rate expression because the eigenvalue of the smallest positive mode at the saddle point $\lambda_2^\ddagger$ approaches zero as shown in Fig. \ref{fig:modes}.a. For the continuum chain model without anharmonic corrections this peak would diverge to infinity as $\lambda_2^\ddagger \rightarrow 0$ at $ N_C$.

Beyond $\tilde N_C$ the saddle point bifurcates into two saddle points where the chain is stretched over the barrier of the external potential as shown in Fig. \ref{fig:modes}.b. There are two saddle points corresponding to the stretched configuration since either one of the two ends can be displaced towards the final state. Since the chain can escape through either of these symmetric saddle points, the total rate is twice the HTST rate given by Eq. \eqref{eq:ratehtst} when $N > \tilde N_C$.

The rate estimate obtained by HTST with AHC saturates to a constant value quickly beyond $\tilde N_C$ as shown in Fig. \ref{fig:rates}. This is because the energy rise along the MEP as well as the positive eigenvalues reach constant values. However, the rate obtained from the molecular dynamics simulations continues to decrease with $N$. The energy along the MEP becomes flat as illustrated in Fig. \ref{fig:meps}.b leading to increased number of recrossings of the TS. When DC obtained from short time classical trajectories started at the TS are included, the rate estimate of Eq. \eqref{eq:ratedc}, close agreement is obtained with the rate obtained from the direct molecular dynamics simulations, the latter being typically a factor of two larger for the longest polymers studied.

\subsection{Langer's Theory}

The escape rate given by Langer's theory of Eq. \eqref{eq:ratelanger} is shown in Fig. \ref{fig:rates} as a function of $N$ and in Fig. \ref{fig:rates} as a function of $K$. For short polymers, below $\tilde N_C$, Langer's rate estimate agrees well with the rate obtained from the molecular dynamics simulations but it underestimates the rate by orders of magnitude for the long chains. The reason for this is the flattening of the energy along the MEP which causes the negative eigenvalue, $\lambda_1^\ddagger$, corresponding to the unstable mode at the saddle point, to approach zero. The Langer-Kramers estimate of the recrossing correction which results in the appearance of $\lambda_1^\ddagger$ in the prefactor in Eq. \eqref{eq:ratelanger} then gives a large overcorrection and much too small rate.

%%%%%%%%%%%%%%%%%%%%
%  DISCUSSION	           		  %
%%%%%%%%%%%%%%%%%%%%

\section{Summary and Discussion} \label{sec:discussion}

The results presented here on the polymer escape problem for the one-dimensional bistable external potential of Lee and Sung \cite{Lee2001} show that a TST approach based on calculations of MEPs to identify saddle points corresponding to the escape transitions followed by a HTST rate estimate with anharmonic and dynamical corrections gives good agreement with results obtained with direct MD simulations for a wide range in polymer length and spring interaction between beads. These TST calculations require little computational effort and can be extended easily to three-dimensional systems with complex interaction potentials, while the MD simulations are limited to short polymers and simple interactions. The computational cost of finding saddle points and computing recrossing trajectories does not depend on the activation energy for the escape. The flat energy profile along the MEP in the diffusion regime makes the dynamical correction trajectories somewhat longer but they are still orders of magnitude shorter than direct MD simulations of the escape transitions. The harmonic mode analysis, furthermore, gives valuable insight into the transition mechanism. The two lowest harmonic modes at the saddle point, shown in Fig. \ref{fig:modes}, explain the relevant escape dynamics of the chain. The lowest, unstable mode corresponds to flux along the MEP towards the final state while the mode corresponding to the second lowest eigenvalue, causes the chain to become stretched for $N>\tilde N_C$, and the saddle point to bifurcate. The eigenvalue of the unstable mode approaches zero as $N$ increases further leading to a flat energy profile along the MEP, as was found by Lee and Sung \cite{Lee2001}. This makes the escape process diffusive along the MEP and the system recrosses the transition state multiple times, calling for explicit dynamical corrections to HTST using short time trajectories started at the TS. The ca. factor of two underestimate of the rate for long chains is likely due to an underestimate of the entropy of the transition state with respect to the entropy of the initial state in the harmonic approximation.  If a full TST calculation were carrier out, followed by a calculation of the dynamical correction, the rate estimate would be exact and agree with the direct dynamical simulations.

The rate estimate obtained from the Kramers-Langer approach which has previously been used extensively \cite{Lee2001, Sebastian2000a, Sebastian2006} turns out to give a poor estimate for the longer chains. The reason is the flat energy profile along the minimum energy path in the diffusion regime which causes the eigenvalue corresponding to the unstable mode at the saddle point to approach zero. As a result, the recrossing correction of the Kramers-Langer approach becomes a large overcorrection and the resulting rate estimate too small by orders of magnitude. Analogous results were obtained by Shin \textit{et al.} using one-dimensional Kramers' theory in the globular limit \cite{Shin2010}. 

The application of HTST to estimate the rate of escape of self-avoiding polymers is more challenging than the ideal polymer escape studied here. The reason is that the initial state where the polymer is coiled up corresponds to multiple local minima on the energy surface. The estimate of the initial state free energy, therefore, needs to take into account multiple configurations and configurational entropy rather than just the energy at a single minimum on the energy surface and vibrational entropy. One way to approach such multiple local minima problems is systematic coarse graining  \cite{Jonsson2011}. Alternatively, the full TST can be applied as opposed to HTST where, for example, the reversible work formulation is used to identify an optimal transition state and estimate the free energy change in going from the initial state to the transition state \cite{Schenter1994,Johannesson2001,Bligaard2005}.

\section*{Acknowledgements} 

This work was supported by the Academy of Finland through the FiDiPro program (H. J. and H.M., grant no. 263294) and the COMP CoE (T. A-N, grant no. 251748). We acknowledge computational resources provided by the Aalto Science-IT project and CSC -- IT Center for Science Ltd in Espoo, Finland.

%%%%%%%%%%%%%%%%%%%%
%  BIBLIOGRAPHY	           		  %
%%%%%%%%%%%%%%%%%%%%

%\end{multicols}

\clearpage

%%%%%%%%%
% FIGURE 1 	%
%%%%%%%%%

%%%%%%%%%
% FIGURE 2 	%
%%%%%%%%%

%%%%%%%%%
% FIGURE 3 	%
%%%%%%%%%

%%%%%%%%%
% FIGURE 4 	%
%%%%%%%%%

%%%%%%%%%
% FIGURE 5 	%
%%%%%%%%%

%%%%%%%%%
% FIGURE 6 	%
%%%%%%%%%

\end{document}